\documentclass[conference]{IEEEtran}

\usepackage[flushleft]{threeparttable}
\usepackage{graphicx}
\usepackage{color}
\usepackage{soul}
\usepackage{todonotes}
\usepackage{hyperref}
\usepackage{amsmath}
\ifCLASSINFOpdf
\else
\fi
\begin{document}

\title{Cloud Kotta: Enabling Secure and Scalable Data Analytics in the Cloud}

\newcommand{\NAMENS} {\textsc{Cloud Kotta}} 
\newcommand{\NAME} {\textsc{Cloud Kotta }}

\author{\IEEEauthorblockN{Yadu N. Babuji, Kyle Chard, Aaron Gerow, Eamon Duede}
\IEEEauthorblockA{Computation Institute, University of Chicago and Argonne National Laboratory\\
\texttt{\{yadunand,chard,gerow,eduede\}@uchicago.edu}
}
}

\maketitle

\begin{abstract}

Distributed communities of researchers rely increasingly on valuable, proprietary, or sensitive datasets. 
Given the growth of such data, especially 
in fields new to data-driven, computationally intensive research like the social sciences and humanities, coupled with what are 
often strict and complex data-use agreements, many research communities now require methods that allow secure, 
scalable and cost-effective storage and analysis. Here we present \NAMENS: a cloud-based data management and analytics 
framework. \NAME delivers an end-to-end solution for coordinating secure
access to large datasets, and an execution model that provides both automated infrastructure scaling
and support for executing analytics near to the data. 
\NAME implements a fine-grained security model ensuring that only 
authorized users may access, analyze, and download protected data. 
It also implements automated methods for acquiring and configuring low-cost storage and compute resources as they 
are needed. 
We present the architecture and implementation of \NAME and demonstrate the advantages it provides in terms of increased performance 
and flexibility. We show that \NAMENS's elastic provisioning model can reduce costs
by up to 16x when compared with statically provisioned models. 

\end{abstract}

\IEEEpeerreviewmaketitle

\section{Introduction}

Data is fast becoming a crucial, if not defining, asset for researchers. 
Entire fields, including those new to computational
practices, are quickly embracing data-driven research. 
However, the increasing scale and complexity of both data and analysis combined with the fact that 
datasets are often proprietary, or sensitive, creates new and unique challenges. 
The emerging centrality of data has led many researchers to design processes around datasets,
which are often housed in tightly coupled environments that discourage reusability and agility. 
To support the needs of data-driven research we have developed \NAMENS\footnote{The name \NAME comes from the Malayalam word for `fortress'.
It represents a secure environment for storing and operating on valuable data.}
, a cloud-based framework that enables the secure, cost-effective management and analysis of large, and potentially sensitive, datasets at virtually any scale.

Many researchers, especially recent adoptees of data-driven practices, lack the infrastructure 
and technical expertise to effectively leverage big data. To satisfy the growing reliance on data-driven research, researchers are increasingly forgoing on-premise infrastructure 
and moving to cloud-based solutions. For example, a variety of valuable and proprietary scientific 
datasets (e.g., 1000 Genomes project and US Census data) are now hosted by Amazon on Amazon Web Services (AWS).

This trend is not difficult to explain: cloud platforms provide high reliability, availability, and performance 
without the need for direct ownership and management of
on-site infrastructure. The adoption of cloud-based services has also facilitated new directions for exploration and investigation. 
For example, researchers can take more risks and more flexibly explore new analyses when storage is co-located with `infinite' elastic computing capacity with which data can be 
analyzed, aggregated, and integrated. 
\NAME aims to catalyze this kind of agility across fluid groups of users (interns, students, postdocs, etc.) while also ensuring 
scalability, security, and data provenance.

While the advantages of big data research on the cloud are evident, they come with unique challenges. For example, storage is 
available with varying performance and cost, identity management can be opaque and complex, and many algorithms used 
to analyze large data are, themselves, complex, computationally expensive, or otherwise unruly. While researchers 
may not concern themselves with the routine management of infrastructure, they are nevertheless, faced with coordinating 
the use of abstract infrastructure. As a result, even cloud platforms have primarily been adopted by large research 
consortia who have the requisite financial resources and technical expertise to manage them.

\NAME addresses research priorities with an analytics environment that allows researchers to 
concurrently run analytics of their own design within a secure environment and in close proximity to the data, with a flexible storage model designed to minimize cost.

\NAME is designed to be accessible to a broad range of users. It is open source and can be used as a service or deployed with automated deployment scripts. 

The remainder of this paper is as follows.
In \S\ref{sec:applications}, we present representative datasets and analyses hosted by \NAMENS. 
In section \S\ref{sec:background} we discuss requirements, before presenting 
the general architecture in \S\ref{sec:architecture} and 
describing the cost-aware mechanisms (\S\ref{sec:cost}), 
and the security fabric (\S\ref{sec:security}).
In \S\ref{sec:evaluation} we evaluate \NAME along a number of dimensions.
Finally, in \S\ref{sec:relatedwork} and \S\ref{sec:summary}, we review related work and summarize our contribution.

\section{Datasets and Analytics}\label{sec:applications}

\NAME is designed to support the hosting and analysis of large datasets. 
To help specify the requirements of \NAME we review representative data and analytics
for which it was designed and is currently being used.

\subsection{Datasets}
\NAME houses a broad range of datasets, including samples of NSF and NIH awards,
patents, as well as large corpora of scholarly publications (\textit{JSTOR}, \textit{ACM}, 
\textit{IEEE}, etc.). Table~\ref{table:datasets}
summarizes the size and properties of several important datasets that
are stored in a mixture of compressed and uncompressed formats (depending on usage requirements). 
Most are stored as file-based raw data, with large collections of metadata residing in a relational database. The sizes of raw data range
from several GB to many TB. 
Each dataset is subject to its own data-use agreement, 
with different access policies defined for various user groups.
In some cases, datasets are publicly accessible (e.g. \textit{Wikipedia}), 
while some are managed for specific research
consortia (e.g., \textit{IEEE} and \textit{ACM}), and others are hosted for specific research groups 
(e.g. UChicago grants). 

\bgroup
\def\arraystretch{1.1} 
\begin{table}
  \centering
  \label{table:datasets}
  \caption{Datasets hosted hosted by \NAME}
  {\footnotesize
    \begin{tabular}{lll}
    \hline
        \textbf{Dataset}                &  \textbf{Data}  & \textbf{Sensitive} \\ \hline
        UChicago Aura Grants            &  \textasciitilde200GB     &  Yes   \\
        Web of Science                  &  \textasciitilde1000GB    &  Yes   \\
        ACM                             &  \textasciitilde16GB      &  Yes   \\
        Annual Reviews                  &  \textasciitilde55GB      &  Yes   \\
        American Physical Society       &  \textasciitilde510GB     &  Yes   \\
        ArXiv                           &  \textasciitilde400GB     &  No    \\
        IEEE                            &  \textasciitilde5500GB    &  Yes   \\
        JSTOR (Journal Storage)         &  \textasciitilde1700GB    &  Yes   \\
        PubMed                          &  \textasciitilde70GB      &  No    \\
        US Patents                      &  \textasciitilde200GB     &  No    \\
		Wikipedia                       &  \textasciitilde20TB      &  No    \\
    \hline
    \end{tabular}
  }
\end{table}
\egroup

\subsection{Analyses}
\NAME aims to host a wide range of analytics algorithms, 
particularly those used in social sciences and humanities, and more specifically, 
those used on the datasets described above. 

Since deployment (3 months at the time of writing), \NAME has been used to execute
a variety of different analyses. Table~\ref{table:analyses} presents requirements from representative
executions conducted using \NAMENS.

\begin{table}[ht]
  \centering
  \label{table:analyses}
  \caption{Analyses conducted with \NAME}
  {\footnotesize
    \begin{tabular}{llllll}
    \hline
        \textbf{Analysis}    & \textbf{Input} 	& \textbf{Nodes} & \multicolumn{2}{c}{\textbf{Per node}} 	& \textbf{Time} \\
														 & Data					&  			& \textbf{Cores} & \textbf{Memory} & \textbf{(hours)} \\ \hline

		LDA                  & 10M (txt)        & 1     & 8  	& 128GB 	& 44  \\
		Word Embeddings      & 10K (txt)        & 1     & 8 	& 250GB 	& 8-10  \\
		Network Analysis     & 20M              & 61    & 64 	& 64GB 	& 264 \\ 
		OCR                  & 10K (pdf)        & 10   	& 32 	& 75GB 		& 20  \\ 
		XML Parsing          & 3.5M (xml)       & 188  	& 1 	& 16GB 	& 3   \\ 
		MF                   & 2.5Kx122         & 1    	& 17 	& 100GB 	& 10  \\ 
		\hline
    \end{tabular}
  }
\end{table}

\subsubsection{Text Analysis}

Text analysis is one of the most common classes of analytics executed by \NAMENS. 
Typically, text analysis is carried out in a multi-stage workflow, where each new stage receives the output of the previous stage.
For example, extracting latent topics from
a publication corpus requires that documents are first normalized and divided into logical bins. Next, a pre-processing stage removes irrelevant text such as common grammatical words, 
proper nouns, and punctuation. The extracted text is then analyzed semantically
using one (or many) models (e.g., doc2vec, word2vec~\cite{Mikolov} and various probabilistic topic models~\cite{Rosen-Zvi,Zhang}) that make sense of the words. 

This process is both memory and compute intensive, relying on
high-performance libraries compiled for specific CPU architectures. 

\NAMENS, has been used to generate a latent Dirichlet Allocation (LDA) model on the Thomson Reuters Web of Science, a large corpus of 10MM documents, using the gensim multicore lda package. This task took 44 hours using 1 instance with 8 cores and 128 GB of RAM.
\NAME has also been used to parse the entire edit history of the English language Wikipedia to explore journal citation practices~\cite{teplitskiy2015amplifying}. Searching over 40TB of XML documents is a highly IO intensive process and requires instances with large (SSD) storage. By processing the 188 data chunks in parallel, \NAME was able to reduce execution time from several weeks to several hours.

\subsubsection{Word Embeddings}
Word embeddings allow researchers to perform discourse analyses on texts. This involves treating documents as sequences of tokens or as `composable' units. This is typically done using structured prediction (with conditional random fields or recurrent neural networks). Jobs of this kind that are run on \NAME involve taking tranches of scientific abstracts, processing each sentence in turn, and training a model to predict where the discursive type (e.g., method, results, discussion, etc.) of the sentence occurs. 
This prediction uses vector-representations of different sections in articles to characterize their relationship to one another. 
Both problems are computationally expensive. 
The training complexity for a conditional random field is quadratic to the size of the label set, and nearly quadratic for the size of the training sample: analyzing 10,000 documents on an instance with 10GB of RAM and 8 cores required 8 hours. 
Training a neural network to produce the word embeddings with 1,000 journals (with varying numbers of articles) required an instance with 250GB of RAM, 8 cores and 8-10 hours of processing time.

\subsubsection{Network Analysis}
Researchers in the computational social sciences are increasingly interested in large scale, complex network analysis. Early work on \NAME was leveraged in developing and analyzing a massive, dynamic hyper-graph model of biomedical science ~\cite{shi2015weaving} and a dynamic hypergraph model of practicing scientists and scholars~\cite{gerow2015proposing}. For the biomedical study, researchers extracted all authors, chemicals, diseases, and methods represented in the National Library of Medicine's 20 million article MEDLINE dataset and constructed a dynamic hypergraph model through time (e.g. 1950 - 2008). Decomposing 20 million records into roughly 9 million authors, 9 thousand chemicals, 4 thousand diseases, and 2 thousand methods and then recomposing these `nodes' into a dynamic hypergraph representation of MEDLINE executed on 61 nodes, each with 64 cores and 64GB of RAM, and ran for 11 days. 

\subsubsection{Optical Character Recognition (OCR) }
OCR software such as \emph{tesseract} is used to extract text, figures, tables, and features from non-text documents, such as PDF.
Given the vast amount of important content locked within non-text documents, OCR is relied upon by many researchers. 
\NAME is used to run OCR software on PDF-based grant proposals and scholarly texts. Extracting text from a corpus of 10K documents using \NAME
required 20 hours utilizing 10 instances with 32 cores and 75GB of RAM. 

\subsubsection{Matrix Factorization (MF)}
When faced with lossy data, researchers use multiple imputation (MI) to recover missing values. Typically, a `missingness' pattern is established on a given response (in a survey, for example) which is used as the dependent variable in a parameterized regression using the non-missing responses as parameters. The `multiple' aspect of MI refers to the process whereby after being imputed, the new values can increase the accuracy of imputing other missing values. This process must be cross-validated for stability and to parameterize error-bounds over sets of multiply imputed data. 
\NAME was used to run low rank and low norm matrix factorization (MF) (an alternative to parametric MI). 
Once the models were developed, \NAME was used to execute a large batch of validations to provide pooled results. 
One such batch, represented as a 2,500 by 122 matrix, consumed 1 instance with 32 cores and 100 GB of RAM for 10 hours.

\section{Requirements}\label{sec:background}

\NAME is designed to address the requirements
of two central use cases: managing community datasets and providing
scalable, analytics capabilities. Here we briefly describe these use cases and 
their requirements.
\\
\noindent
\textbf{Managing community datasets}. There is a growing need to make valuable datasets
available to research communities. Often, this requirement is motivated
by funding agencies or institutions. However, it is also a proven means 
for establishing and growing research communities around shared datasets. 

The requirements for this use case are that \NAME be:

\begin{itemize}
	\item Secure: data must be securely stored and accessible only to
	authorized users. 
	\item Scalable: storage must scale
	to meet the needs of increasingly large datasets and access workloads. 
	\item Reliable: data must be stored reliably, with efforts made to 
	backup data in case of failure, corruption, or disaster. 
	\item Available: data must be available to a broad set of 
	geographically distributed users with minimal downtime.
	\item Cost-effective: the costs associated with high performance, secure, 
	reliable, and available data storage must be relatively low. 
	\item Analyzable: data value is most often derived from analysis. 
	Data should be easily and efficiently analyzed with various tools.
\end{itemize}

\noindent
\textbf{Scalable analytics}.

As data sizes grow and analysis
algorithms become more computationally intensive, the required resources often exceed
those available to researchers. Thus, methods are required to scale analyses from individual
computers to distributed and parallel computing systems.
The requirements for this use case are that \NAME be:

\begin{itemize}
	\item Secure: authorizations control what data can be analyzed and 
	users' analyses must be isolated from one another. 
	\item Scalable: analyses must scale to the size of data, exploit
	parallelism where possible, and leverage large scale computing infrastructure
	for efficient performance. 
	\item Cost-effective: analysis costs must be comparable to, or lower than, 
	that of using local compute resources.
	\item Easy to use: interfaces must minimize the complexity of using the 
	underlying infrastructure. 
	\item Co-located with data: data movement can be costly and / or impose
	significant overheads. Where possible, analytics workloads
	should be placed to minimize data transfer.
\end{itemize}

\section{Architecture and Implementation}
\label{sec:architecture}

\begin{figure}
  \center
  \includegraphics[width=0.45\textwidth]{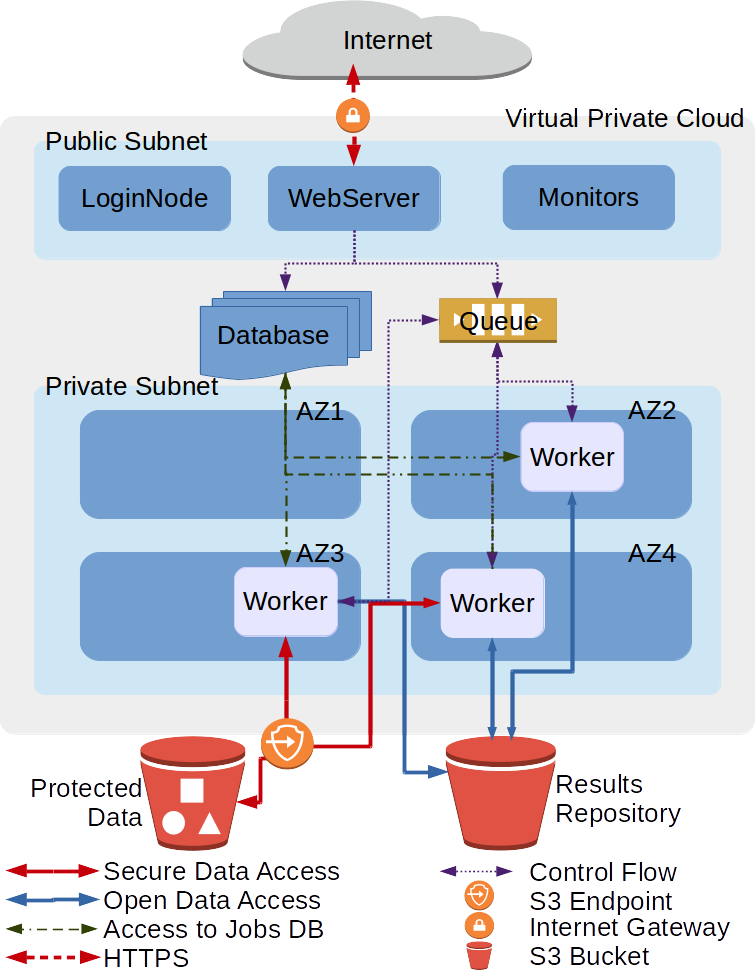}
  \caption{Architecture of \NAMENS.}
     \label{fig:arch}
    \vspace{-1.5em}
\end{figure}

The \NAME architecture is depicted in \figurename~\ref{fig:arch}.
The entire system is comprised of a web interface and web service;
a storage layer that provides fast, reliable, cost-effective storage;
a compute layer that provides elastic and cost-effective compute resources;
a job management layer that provides reliable execution of user-specified jobs; and 
a security fabric that permeates the whole system. 
It also includes a collection of automated deployment and configuration scripts,
as well as monitoring and management software. 

\NAME is designed to be deployed on \textit{Amazon Web Services} (AWS), the research ecosystem of its intended users.
Where possible, \NAME builds upon existing cloud services as they are 
scalable, reliable, secure, and cost-effective.
The entire \NAME system is open source and can be deployed using a reproducible \textit{CloudFormation} configuration.

\subsection{User Interface}
\label{sec_interface}
\NAME offers three interfaces: 
a web interface, a REST API, and a command line interface (CLI) accessible from the login node. 
This range of interfaces supports broad usage scenarios, enabling intuitive web access
for web-based users alongside advanced programmatic and CLI support to facilitate customizable
and automated invocation. 

The supported interfaces allow users to browse datasets, upload new data, view and download
results from previous analyses, and submit and manage new analyses.

All interfaces are secured, restricting access to only properly
authenticated and authorized users. 

Users can upload files to their own private storage, and browse accessible data.
Once uploaded, files are available to be specified as inputs to submitted jobs.

In \NAMENS, a job consists of a complete description of an executable, a list of inputs,
a list of output files to be saved, a maximum wall-time, and a target queue.
In addition to supporting arbitrary executables, applications can be templated to create 
pipelines with simplified user interfaces.
Users submit jobs via simple web forms, or by specifying the job as a JSON file for
the CLI and REST interfaces.

\subsection{Storage Layer}
At the heart of \NAME is a durable storage layer that provides scalable storage of managed datasets. 
The storage layer uses several AWS services that provide different
guarantees regarding access time, durability, and availability with different cost 
models. The types of storage used by \NAME are:

\begin{itemize}
  \item \textbf{Elastic Block Store (EBS)}: a high performance block storage model that can be mounted as a file system. 
  \item \textbf{Simple Storage Service (S3)}: a reliable object store that provides high performance access via HTTP(S).
  \item \textbf{S3 \textit{infrequent access}}: an object store with reduced storage cost at the expense of increased data access cost.
  \item \textbf{Glacier}: an archival storage model that provides high durability at a low price with high data retrieval times.
\end{itemize}

Rather than rely on a single storage tier for all data, \NAME implements a data lifecycle model where
data are migrated between storage tiers based on access workloads.

\subsection{Compute Layer}

The workloads for which \NAME is designed often comprise independent, loosely coupled jobs. 
As such, they are well suited for execution in a high throughput computing model. 
To address these needs, \NAME implements a scalable compute 
layer based on an elastic pool of AWS Elastic Compute Cloud (EC2) instances.

EC2 offers a range of different instance types (virtual machines with fixed resources). 
Instances are organized by region and \textit{Availability Zone} (AZ).
Regions represent different geographic locations whereas 
AZs are located in a specific region and offer independent failure probabilities. 
EC2 instances are provisioned according to a market model in which users
pay for the resources consumed. \NAME can be configured to use two different EC2 market models: 

\begin{itemize}
  \item On-demand instances are offered at a fixed hourly price.
	There is no long-term commitment and an instance will remain
	operational until it is terminated by the user. 
  \item Spot instances are offered using a dynamic price model 
	where users specify the maximum hourly price they are willing to pay 
	and instances are provisioned until their price exceeds the user's bid. 
	Spot instances tend to be a fraction of the price of their
	on-demand equivalents, but they may be terminated without warning. 
	\end{itemize}

Like HPC systems, \NAME is used for two distinct types of workloads: short development tasks requiring quick responses but minimal compute resources, and longer running production tasks that are computationally intensive but more tolerant to delays.

Given the nature of our target workloads (many independent jobs) we adopt a queue 
model and implement two logically independent pools. To guarantee that development jobs do not wait for long periods of time, the development pool is 
always provisioned with at least one reliable (on-demand) instance.
In contrast, the production queue uses Spot instances to reduce costs.
\NAME provisions additional instances when there are pending jobs in the queues.

\subsection{Job Management}

When submitting an analytics job users define a task description that includes
the analysis script, the required inputs and which output files 
to save to persistent storage, and other configuration settings.
Upon submission, the entire description is stored in the database.
The job management system adds the user's role identifier to the task description and forwards 
it to the appropriate queue for execution.

The queue provides a reliable way of managing task execution. 
Worker nodes, when first instantiated or idle, poll the queue for waiting tasks. 
The worker retrieves the queued job, looks up the job description in the database, 
and starts executing the job. 
Because \NAME makes use of Spot instances, failures stemming from instance revocation are 
not uncommon. A queue-watcher service monitors nodes for early termination (or other failures)
and resubmits tasks to the queue in the case of failure.
Throughout execution the worker node writes job status markers to the database. 
This provides a constant stream of worker statistics (CPU, I/O and RAM utilization) and job progress which
is can be interrogated via the web interface to provide real-time feedback.
When the job completes, output data is staged to the user's storage,
the completion code of the application is written to the database, and the
worker node marks itself as idle to begin the queue polling process.

\section{Automated Cost-Aware Mechanisms}\label{sec:cost}

\NAME is differentiated from comparable systems via its use of automated, policy-based
mechanisms to reduce costs and improve performance of storage and compute.

\subsection{Storage} \label{subsec:storage_cost}
\begin{figure}
  \center
  \includegraphics[width=0.45\textwidth]{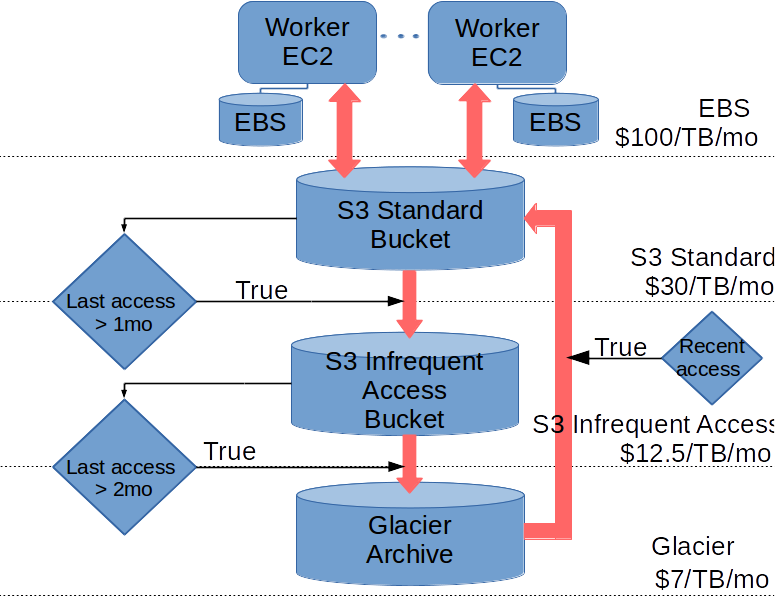}
  \caption{Storage tiers in \NAME and the heuristics used to minimize storage costs. Some costs may differ across regions and configurations.}
    \label{fig:storage}
    \vspace{-1.5em}
\end{figure}

\NAMENS's storage layer incorporates various storage tiers with different properties. 
An automated data lifecycle model manages data across tiers by applying a Least Recently Used (LRU) 
caching strategy to data (\figurename~\ref{fig:storage}).
The primary store for data is S3.  
When data is needed for analysis, it is either retrieved directly from S3 or 
it is staged from another tier via S3. 
Data is made available to a job via ephemeral storage on the instance
or an attached EBS volume. 
At the conclusion of analysis, output files are staged back to S3. 

\NAME uses S3-Standard (STD) and S3-Infrequent Access (IA) tiers for frequently accessed datasets 
and Glacier's low cost storage for less frequently accessed data.  
\NAME can be configured with a LRU \emph{staleness} property that defines how long data is 
stored in a particular tier. For example, the policy \emph{STD30-IA60-Glacier}
will move data from STD to IA if it is not accessed for 30 days, and from IA
to Glacier if it is not accessed for a further 60 days. 

When data is stored in Glacier there is potential for significant delays accessing data.
If analyses are submitted requiring data that is stored in Glacier, the job management
system will identify that the data is not available and submit a 
request for it to be retrieved from Glacier. The analysis job is placed 
in a separate queue until the data is available in S3. When the data is available
the job will execute as normal.

\NAMENS's storage model has important advantages over a static storage configuration.
First, while EBS provides low-latency access, it must be mounted as a file system to 
access data and its persistent nature can result in significant costs. 
By storing data in S3, a small overhead is incurred to stage
data for compute, however, this latency is nominal in most cases, representing a fraction of the total time it takes to provision, and execute a job.
Finally, Glacier provides low cost and reliable storage if reduced availability can be tolerated.

\subsection{Compute}

Most production jobs are computationally intensive, long running tasks, that are tolerant of delays.
To minimize costs, \NAME aims to host production jobs on Spot instances where possible.

\NAME uses an automated provisioning model to acquire the instances needed to execute a job.
It is able to provision resources across all AZs in a region to minimize the impact of price spikes local to a single AZ. 
Administrators can define static or policy-based bid prices (some fraction of the equivalent on-demand price, for example). 

While Spot instances can significantly reduce costs, instance revocations are inevitable. 
In this case, \NAME can reschedule running jobs on a different Spot instance
to ensure the job will complete, albeit with an increased execution time. 

There are a number of trade-offs that must be considered regarding compute cost, execution time, 
and wait time. To provide flexible control of these trade-offs, \NAME offers various policy-based
configuration options. For example, administrators may
set maximum bid prices, define the minimum and maximum number of instances per pool, and 
select suitable instance types for execution. Presently, \NAME uses a single pre-selected instance type 
for the development and production pools. When provisioning Spot instances for the production pool
the cheapest instance across AZs is selected by default. 
In future work, we intend to integrate cost-aware provisioning~\cite{chard2015costworkloads, chard2015cost}
and profiling~\cite{chard2016profiling} models to automate the selection of 
instance types based on an analysis of predicted cost and execution time.

\section{Security}\label{sec:security}

\begin{figure}
  \center
  \includegraphics[width=0.45\textwidth]{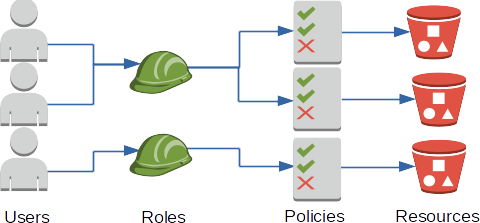}
  \caption{Role-based security model in \NAMENS.}
    \label{fig:security}
    \vspace{-1.5em}
\end{figure}

\NAME implements a flexible and extensible, role-based access control
model across all resources managed by the service.
As shown in \figurename~\ref{fig:security}, users are assigned roles from a list of predefined roles, for example \emph{kotta-public-only} and \emph{kotta-read-WOS-private}, 
where \emph{WOS} refers to access to the private Web of Science dataset.
Policies define a role's privileges on a specific resource.
All data access is controlled by user roles and, as such, worker nodes
must assume a user role before being able to access restricted data.
Internal services, such as the queue watcher, are granted appropriate privileges by internal roles 
such as \emph{web-server} or \emph{task-executor}.
These roles, unlike user roles, have access to the internal database, queues and notification systems and are capable of controlling scaling functionality.

Since \NAME uses \emph{Login with Amazon} for authentication, users require
an account with Amazon's merchant service to use \NAMENS.
Before being granted permission to use the system, the user's 
unique identity must be registered and mapped to a role.
Users may then authenticate using Amazon's OAuth~2 interface. 
Following the redirection-based OAuth~2 workflow, the user is redirected
to a secure AWS website to authenticate. 
Upon successful authentication, a short-term
delegated access token is returned to \NAMENS. The token is valid for one 
hour and during that time it can be used to perform actions on behalf of the
authenticated user. 
In keeping with the principle of least privilege, every user in \NAME starts with 
no privileges and is incrementally granted permissions when required.
This strategy can increase the burden on system administrators who are responsible for managing policies and roles. Nevertheless, it ensures that the system remains secure which is a priority for users and administrators.

The \NAME web interface translates access tokens into short duration 
web sessions (using cookies) valid for six hours. 

All data access control is implemented at the S3 level. 
Every S3 bucket may have policies which prescribe role-based access permissions. 
For private data not available for download, policies are configured to allow for
read-only access to specified compute nodes. 
For datasets that can be downloaded, 
policies restrict access to authorized users.
The data stored on S3 buckets are server-side encrypted and accessible only from a Virtual Private Cloud (VPC) Endpoint. 
This guarantees
that traffic between S3 buckets and the compute instances remain private.
Derived data as the result of an analysis is stored as private objects which
can be managed only by the creator. \NAME also supports a short-term
signed URL model (similar to sharing links in DropBox) for securely sharing data. 

The compute layer is insulated from the Internet by a private subnet enclosed in a VPC.
Worker nodes in the compute layer are provisioned with a specific \emph{task-executor}
role that grants them a minimal set of privileges
such as read access to the database, queues, and to the S3 bucket where data can be accessed and results can be stored.
Most importantly, this role is a trusted role that is authorized to allow for switching between other user roles.
When a worker receives a task, it switches to the role of the user to stage input data.
As a result, objects in S3 buckets can be accessed from worker nodes provided that the user's 
role is authorized to access them.

After input files are staged, the worker resumes its \emph{task-executor} role and 
continues execution of the job. During execution,
temporary credentials are used to
record progress in the database and store intermediary and output data. 
After tasks terminate, any output files are transferred to S3.

An important component of \NAME is the ability to audit data usage.
To do so, \NAME tracks all data access by users and analyses, which
is combined in audit logs. 

\section{Evaluation}\label{sec:evaluation}
Our evaluation explores several important aspects of \NAMENS.
First, we investigate production usage of the system.
Second, we explore the benefits of storage lifecycle policies. 
Third, we interrogate elastic resource provisioning
with regards to cost and makespan.
Fourth, we evaluate the throughput of the system using
a worst-case, many small jobs workload.
Finally, we examine the potential benefits of
cost-aware provisioning when considering the cost of data transfer.

\subsection{Production Usage}
\begin{figure}
  \center
  \includegraphics[width=0.475\textwidth]{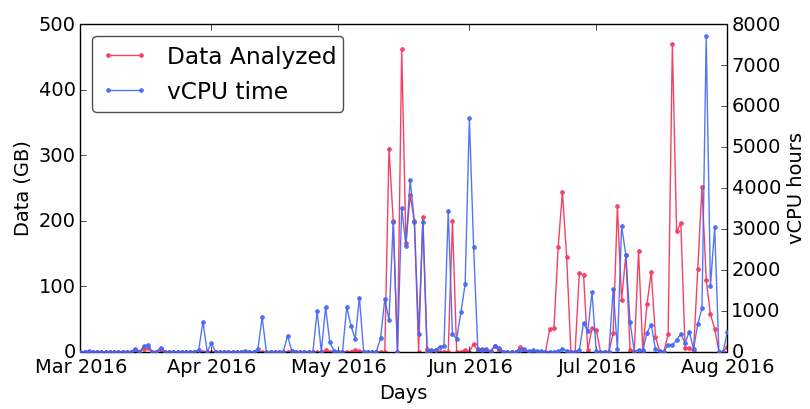}
  \caption{System utilization}
  \label{fig:usage}
  \vspace{-1.5em}
\end{figure}

\NAME has been used in production by a number of researchers for a diverse
range of applications. \figurename~\ref{fig:usage}
shows the total data analyzed and the total number of compute hours used, per day over the last 3 months. 
Researchers have used \NAME to process more than 5TB of data with over 75,330 CPU hours.
Individual days approach 500GB of data analyzed with nearly 8,000 CPU hours.
The figure shows that compute is generally proportional to data size, with
several exceptions. Specifically, the differences in late June and August are
due to long running machine learning models and text processing on Wikipedia datasets.
The graph also highlights the sporadic usage of our users, which reinforces
the value of elastically provisioning infrastructure when required.

\subsection{Storage cost evaluation}

To evaluate our adaptive storage model we consider each of the 
storage services provided by AWS in isolation as well as in combination when lifecycle policies are used.
We assumed a fixed dataset of 10TB and calculate storage costs with two access workloads.
We calculate storage costs for each tier based on advertised pricing. 
S3 STD and IA are offered at a tiered cost per byte. Note that we do not include S3 data access costs as they are negligible (\$0.004 per 10,000 requests within AWS).
Glacier is offered at a fixed cost per byte with an additional retrieval fee for accessing archived objects (5\% of average monthly storage can be retrieved for free).

We model the costs of storage in Glacier as follows.
The cost per month of storage ($C_{mo}$) depends on the peak transfer rate $Tx_{p}$ calculated from the peak daily transfer volume $D_{daily}$ (assumed to be retrieved in $Tx_{time} = 4hours$).
There is a cost for transferring data from Glacier $C_{tx}$ if transfer volumes exceed the daily pro-rated transfer quota $Q_{tx}$ as a percentage of all data stored in Glacier $D_{glacier}$. We model the monthly cost by scaling the transfer ratio over 30 days.

\setlength{\belowdisplayskip}{0pt} \setlength{\belowdisplayshortskip}{0pt}
\setlength{\abovedisplayskip}{0pt} \setlength{\abovedisplayshortskip}{0pt}

\begin{equation}
\label{eqn:3}
    Tx_{p} = {\frac{Dx_{daily}}{Tx_{time}}},   Tx_{q} = {\frac{D_{glacier} \cdot 5\% }{30 \cdot {Tx_{time}}}}
\end{equation}

\setlength{\belowdisplayskip}{0pt} \setlength{\belowdisplayshortskip}{0pt}
\setlength{\abovedisplayskip}{0pt} \setlength{\abovedisplayshortskip}{0pt}

\begin{equation}
\label{eqn:9}
 C_{mo} =
\begin{cases}
    0 ,& \text{if } Tx_{p} < Tx_{q}\\
    (Tx_{p} - Tx_{q}) \cdot C_{tx} \cdot 720 , & \text{otherwise}
\end{cases}
\end{equation}

Analysis of production data access in \NAME indicates that only a small fraction $A_{data}$ (3-10\%) of the total data is accessed in a 3 month period.
When applying a lifecycle policy (e.g., STD30-IA60-Glacier), the monthly storage cost $SCo_{mo}$ is then:

\begin{equation}
\label{eqn:11}
     SCo_{mo} =  {\frac{(C_{std}+2C_{IA})}{3}}(1-A_{data})+(C_{glacier} \cdot A_{data})
\end{equation}

Table~\ref{table:storage_cost_eval} shows the cost for storing and accessing data in S3 STD, S3 IA, Glacier, and two lifecycle policies with different data access ($A_{data}$) rates (3\% and 10\%).
The results show that storage costs can be significantly reduced
by using S3 IA and Glacier. However, access costs and the time to retrieve data when using Glacier
may negate these benefits. Our storage lifecylce policies are able to balance these costs by automatically
moving data between storage classes.

\bgroup
\begin{table}
  \begin{threeparttable}
  \centering
  \label{table:storage_cost_eval}
  \caption{Storage Cost Projection for 10TB over a year}
  {\footnotesize
    \begin{tabular}{llll}
      \textbf{Storage Strategy}   & \textbf{Cost} & \textbf{Access cost} & \textbf{Access time}\\ \hline
      S3-Standard		         			& \$3546	      & NIL       & NIL    \\
      S3-Infrequent Access 	  	  & \$1500        & NIL       & NIL    \\
      Glacier (3\%)               &	\$840         & \$4217.2  & 4hours\tnote{\textdagger}\\
      STD30-IA                    & \$1670.5      & NIL       & NIL    \\
      STD30-IA60-Glacier (3\%)         & \$880.259     & \$169.73  & 4hours\tnote{\textdagger}\\
      STD30-IA60-Glacier (10\%)          & \$974.20     & \$169.73  & 4hours\tnote{\textdagger} \\ \hline
    \end{tabular}
  }
  {\begin{tablenotes}
      \item[\textdagger] :Average glacier retrieval time
  \end{tablenotes}
    }
\end{threeparttable}
\end{table}
\egroup

\subsection{Elastic Scaling}

\begin{figure*}[ht]
  \center
  \includegraphics[width=\textwidth, height=8cm]{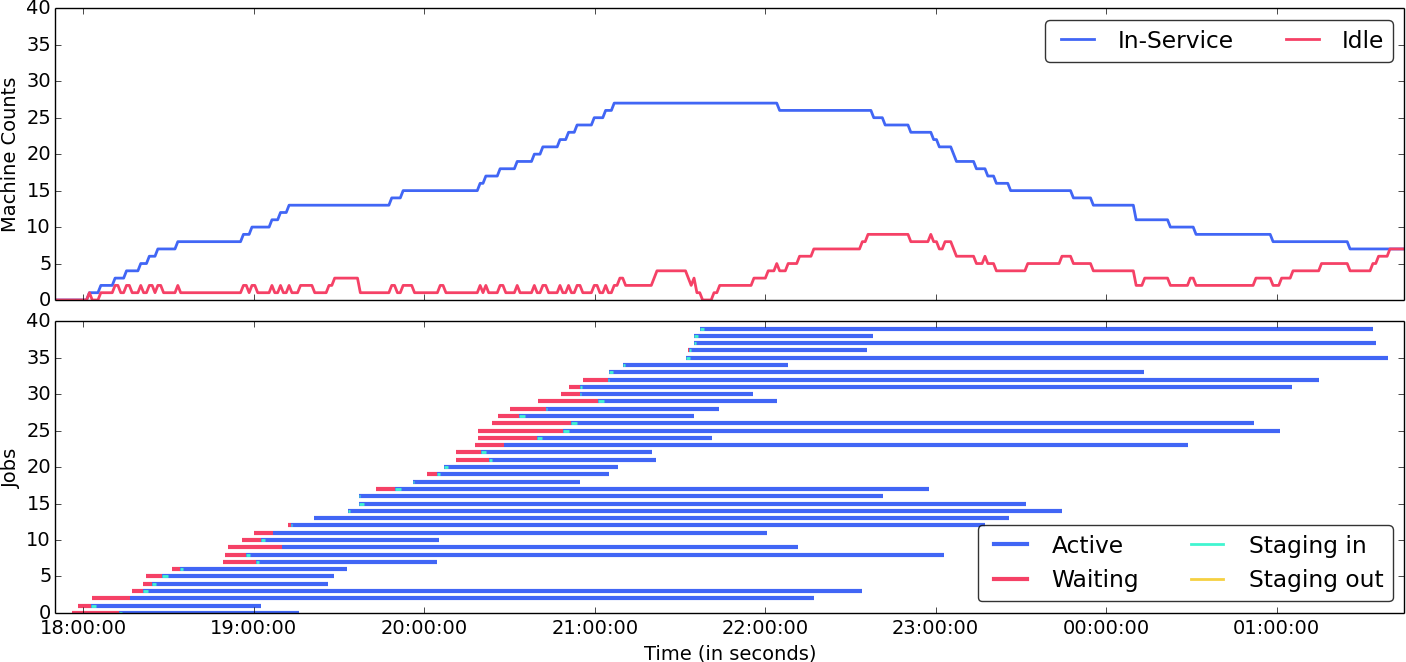}
  \caption{Elastic scaling in \NAMENS. The top plot shows currently provisioned nodes and the proportion of which are idle. The bottom plot shows each job's status from submission to completion, including wait and data staging time.}
  \vspace{-1.5em}
  \label{fig:exp12}
\end{figure*}

\NAME embraces on-demand and elastic cloud computing capacity by dynamically provisioning
new instances to host submitted workloads.
Rather than relying on a time-sharing system or scheduler, scaling is achieved by provisioning instances as the need arises based on the state of the queue.

For each experiment we compared the trade-offs between total execution time (`makespan') and total cost.

Three scaling strategies were explored:
\begin{itemize}
    \item \textbf{No scaling}: A baseline strategy in which a fixed number of instances are provisioned.
    \item \textbf{Limited scaling}: A restricted strategy where the maximum number of provisioned nodes is limited.

    \item \textbf{Unlimited scaling}: An unbound strategy where as many instances as are needed will be provisioned. 
\end{itemize}

To evaluate elastic scaling under a realistic usage scenario we created a simulation
workload to mimic existing production
usage of \NAME (see Section~\ref{sec:applications}).
To reduce costs of experimentation, the workload
consists of 40 jobs submitted over a four hour period.
The inter-arrival time is obtained from a Poisson distribution
of the form $t \sim \mathcal{P}(\lambda=0.1667)$ hours.
Where $\lambda$ was selected based on the duration of the experiment.
Within the workload, we modeled three distinct job types representative of three distinct analyses that have been executed on \NAMENS.
Specifically, jobs were configured to run for
1, 3, and 4 hours, with 40\%, 20\% and 40\% in each category, respectively.
Each job duration was further varied by up to $\pm5\%$ minutes to
ensure the results are not biased toward hourly increments.
To model data transfer time we randomly assigned input datasets of
$\{1,3,5,7,9\}$ GB. These datasets were hosted on and staged from S3. 
The jobs themselves were comprised of simple calls to \texttt{sleep()}.

\figurename~\ref{fig:exp12} illustrates the relationship between job execution
and the elastic infrastructure provisioned to host the workload when using the unlimited
scaling strategy.

For each job, it shows
the time of submission, the wait time in a queue before execution began,
the data staging time (in/out), and the execution time.

The results highlight \NAMENS's ability to elastically provision infrastructure
for waiting jobs. The workload peaks at 27 concurrent jobs.
The results also show a moderate wait time due to the delay caused by
provisioning instances, which, on average, is 7:39 per job with a 
peak of 30 minutes due to spot market volatility.
The advantage of reusing existing instances is made evident in the figure. For instance,
the last 7 jobs were able to execute without waiting because idle instances were
available in the pool.

Table~\ref{table:provisioning} compares the different scaling strategies with respect
to makespan, cost, and wait time.
Wait time is the time a job waited in the queue.
Makespan is the total execution time from when the first job is submitted until the last
job completes.
Spot costs were calculated as the cost paid for running the experiment
using Spot instances.
The spot costs may vary significantly because these experiments were run at different times with
different market conditions between executions.
The on-demand cost was calculated based on the price that would have been
paid had on-demand instances been used. These
results aim to remove the variability of the spot market when comparing costs.
The cost savings, then, are the percentage improvement
compared to the baseline (no scaling, min: 40, max: 40) strategy.

The no scaling strategy, where a fixed pool of instances are available to host
the workload, represents a baseline strategy to which the others can
be compared. Using a fixed pool of 40 instances, the wait time is 0, thereby fully optimizing
the makespan. However, many instances are idle for a significant portion
of the workload which results in costs of \$74.57 for 40 fixed instances
and \$40.87 for 20 fixed instances.
The unlimited scaling strategy provisions instances only when they are required.
Here the wait time is longer than the no scaling strategy (on average 7:39
per job), however the cost of executing the whole workload is significantly lower
(saving approximately 61\% using spot or on-demand instances).
The unlimited scaling strategy using Spot instances offers similar compute performance 
at \( \frac{1}{16} \) the cost of a static cluster provisioned with on-demand instances.
It is worth noting that the makespan is the same here given that the last jobs to complete do not have to 
wait in the queue due to available idle instances.
The limited scaling strategies aim to provide a hybrid model to manage
the trade-off between cost and time. As expected, the results show
increased makespan (roughly 1 hour and 5 hours longer for 20 and 10 instances, respectively)
with reduced cost (roughly \$2 and \$5 for 20 and 10 instances, respectively).
Our results show, that knowledge of workloads can be used to define 
scaling restrictions that minimize idle times and reduce costs.

\bgroup
\def\arraystretch{1.2} 
\begin{table*}
  \centering
  \label{table:provisioning}
  \caption{Cost vs. Makespan.}
  {\footnotesize
    \begin{tabular}{lllllcccc}
    \hline

\textbf{Scaling} & \textbf{Nodes}     &                   & \multicolumn{2}{c}{\hspace{-.7cm}\textbf{Cost}} &\multicolumn{2}{c}{\textbf{								Wait Time}}&   \multicolumn{2}{c}{\% \textbf{Savings}} \\
							   & \textbf{(min,max)} & \textbf{Makespan} & \textbf{Spot}        & \textbf{On-demand}       &  \textbf{Max}          & \textbf{Avg.}                 &    \textbf{Spot}  & \textbf{On-demand}   \\

 \hline
None		&	40,40  & 07:43:00 & \$10.26   & \$74.57   & 00:00:00      & 00:00:00    &  0       &   0        \\
None 		&	20,20  & 08:33:00 & \$5.98    & \$40.87   & 01:27:00      & 00:11:30    &  41.71   &   45.19    \\
Unlimited   &	0,-    & 07:43:00 & \$3.95    & \$28.92   & 00:30:00      & 00:07:39    &  61.50   &   61.21    \\
Limited 	&	0,20   & 08:22:00 & \$4.52    & \$26.77   & 01:46:00      & 00:15:10    &  55.94   &   64.10    \\
Limited		&	0,10   & 12:50:00 & \$3.62    & \$23.18   & 05:41:00      & 02:08:06    &  64.71   &   68.91    \\

    \hline
    \end{tabular}
  }
\end{table*}
\egroup

\subsection{Throughput}
To evaluate the throughput of \NAMENS, we designed a strong scaling experiment where 10,000
small tasks were submitted to a pool of general purpose instances (m4.xlarge, 4 cores@2.4Ghz).
To model a worst-case scenario, each task is a \texttt{sleep(0)} call and requires
no data staging.
Instances were provisioned ahead of time to reduce overhead associated with the provisioning process.
We measure the total time to completion of the 10K tasks for $\{1,2,4,8,16,32\}$ worker nodes.
In an ideal case, we should observe speed-ups that are directly proportional to the number of nodes.

\begin{figure}
  \center
  \includegraphics[width=0.475\textwidth]{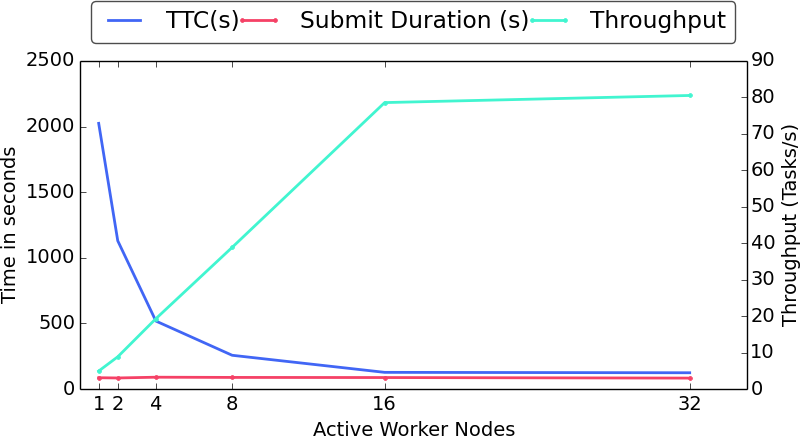}
  \caption{Throughput in \NAME for 10,000 tasks}
  \label{fig:throughput}
  \vspace{-1.5em}
\end{figure}

\figurename~\ref{fig:throughput} shows the time to submit all 10,000 tasks, the total time to completion
for tasks on provisioned instances, and the throughput.
The results show that throughput increases linearly with the number of worker nodes
up to 16 nodes. Up to this point, the average task throughput per worker node is 4.90 tasks/s (total 79.84 tasks/s).
In this experiments, the primary bottleneck is the database as transactions are used to record job descriptions and
performance markers. So, for these experiments, we increased the DynamoDB read and write capacity per second to 100 and 400, respectively.
It is important to note that these experiments represent an idealized worst case
scenario in which tasks are trivial and frivolous. With more typical, longer running jobs that run on the order of
minutes or hours, our results show that \NAME can easily support thousands
of jobs submitted simultaneously.

\subsection{Cost-aware Provisioning}

Here we explore the benefits of various provisioning strategies in the context of geographically distributed computation in a cloud environment.
In this experiment, the data was hosted in one region (N.Virginia, us-east-1) and compute was distributed across different EC2 regions and AZs.
We use historical Spot price data from ten AZs spread across four regions, and simulate various strategies for instance selection.
We simulate a single hour task that is provisioned among the AZs based on the different strategies.
This simulation is executed for a month to capture the long term impact of price differences between AZs.

We calculate the total cost of computation for a given hour $P_{total}$
as the sum of the hourly instance price $P_{i}$ and the costs
associated with data transfer.
\begin{equation}
\label{eqn:1}
    P_{total} = P_i + P_{transfer}
\end{equation}
If the selected compute instance lies in a different region than the location of the
S3 data bucket, there is an inter-region data transfer cost $T_{c} = \$0.020/GB$ ~\cite{s3_transfer_cost}.
The amount of data downloaded and uploaded from S3 are represented by $D_{dn}$ and $D_{up}$, respectively,
which we, further, assume to be equal.

\begin{equation}
\label{eqn:2}
 P_{transfer} =
\begin{cases}
    0 ,& \text{if } us-east-1\\
    (D_{dn} + D_{up}) \cdot T_c, & \text{otherwise}
\end{cases}
\end{equation}

\begin{figure}
  \center
  \includegraphics[width=0.475\textwidth]{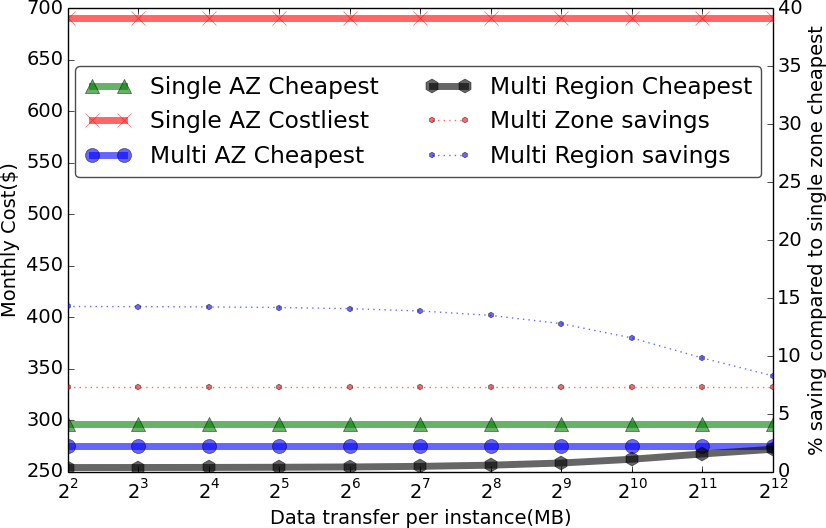}
  \caption{Monthly cost for C4.8xlarge instance with data egress costs.}
  \label{fig:cost_data_egress}
  \vspace{-1.5em}
\end{figure}

\figurename~\ref{fig:cost_data_egress} shows the total monthly cost of using C4.8xlarge
instances when different amounts of data are used by the analysis. 
The graph shows several different provisioning strategies:
selecting the cheapest or most expensive instance in a single AZ,
the cheapest instance within regions and across regions.
The graph also shows the savings that are obtainable when broadening the search scope across AZs and regions.
We observe a significant difference between the cheapest and most expensive
instance within an AZ. This suggests that there is considerable financial risk when provisioning
instances within a single AZ. Thus, there is a considerable advantage to
selecting instances across AZs and regions. Interestingly, however, as the volume of data transfer associated with compute tasks increases, we observe diminishing returns from
extending the search scope to consider multiple regions.
This result confirms the need to co-locate compute with data as data sizes increase.

\section{Related Work}\label{sec:relatedwork}

Currently, there are no existing frameworks that
enable secure and scalable storage, dissemination, and analysis of 
research data using cost-effective, elastic cloud resources. 
Here, we review related work that has significant commonalities with \NAMENS.

In the social sciences there is a growing need to provide secure 
storage and analysis of data~\cite{borner2011plug,babuji2016secure}. While there is yet to be a
complete solution, there are several efforts that overlap with
the goals of \NAMENS. For example, hybrid cloud models have been 
used to support scalable social science
analytics in research computing centers~\cite{abramson14hybrid}. 
Others have extended common software used by social sciences, 
such as Microsoft Excel, to analyze increasingly large data~\cite{saleem14bigexcel}. 
However, these efforts focus primarily on moderately sized tabular data
and interactive analytics. 
The data capsule~\cite{zheng14capsules} model used by the 
Hathi Trust enables secure, non-consumptive analysis of data by leveraging 
controlled virtual machines.

There are many community hosted data repositories designed
to support specific research communities. For example,
dbGaP (database of Genotypes and Phenotypes)~\cite{dbgap} organizes the
results of studies of genotype and phenotype interactions and
NOAA's National Climatic Data Center~\cite{ncdc} provides public access
to national climate and weather data.

Each of these data storage repositories requires significant administrative overhead
to populate, curate, operate, and manage.

In many cases, proprietary authentication and authorization
frameworks have been developed to control access to data.
Though, more importantly, these systems act as 
static, isolated data environments, and provide only minimal
data management capabilities with no computational capabilities.

The Integrated Rule-Oriented Data System (iRODS)~\cite{irods} provides policy-based, federated data management. 
Among other features, it allows distributed file systems to be integrated, 
data to be organized in global namespaces, and rule specification for managing 
data throughout its life-cycle.
However, iRODS is designed to manage file systems, and does not support cloud storage models. 

There are many other systems that provide the ability to deploy cloud-based clusters. 
For example, CloudMan~\cite{cloudman} and StarCluster~\cite{starcluster}
enable deployment of fully functional clusters
for hosting and executing workflows.
Systems like these and others, are primarily designed to aid in the creation
of clusters for semi-permanent usage.
Other systems, such as the Globus Galaxies platform~\cite{madduri2014globus}
and Makeflow~\cite{albrecht12makeflow}, enable on-demand and elastic cluster  
provisioning in response to user submitted workload. 
\NAME is unique, however, in
its use of commodity AWS services and its broad focus 
on providing a framework for secure data storage and analysis.

Science gateways~\cite{wikinsdiehr07gateways} are designed to abstract
the technical challenges that come with using large scale computing
infrastructure.
They typically provide access to shared datasets and resources through 
high level user interfaces (e.g., workflows and portals). 
Examples of commonly used gateways include CyberGIS~\cite{liu13cybergis} for
geoscience and iPlant~\cite{stanzione11iplant} for ecology.
While there is increasing interest in cloud-based solutions, most science gateways are built on more traditional High Performance Computing (HPC) infrastructure~\cite{wu11cloud, madduri2014globus}.
\NAME acts as a fabric on top of which cloud-hosted
gateways could be developed in a domain-agnostic setting. 

\section{Summary}
\label{sec:summary}

\NAME enables distributed groups of researchers to manage valuable and large-scale research 
data and execute complex, heterogeneous analyses in a secure, scalable manner. \NAME helps 
users interact seamlessly with secure datasets while simultaneously enabling administrators to consolidate and 
simplify storage and computational infrastructure, all the while significantly reducing costs. 
Its automated storage lifecycle model allows for these reductions with minimal effect on active research. 
The scaling computing model, which leverages
elastic, low-cost compute resources, further ensures that compute resources are used efficiently and that compute costs are minimized.
By integrating existing identity management tools, \NAME allows for
role-based management of datasets that ensures compliance with a group's range of data-use 
agreements. It is true that any middleware platform comes at the cost of disruption to users' workflows. 
\NAMENS, however, minimizes this disruption by ensuring that users have full privileges on compute nodes 
so they can configure their environment to meet the needs of their analyses. 
Lastly, \NAME implements intuitive interfaces that support users with varied technical competencies.
As computational and data science becomes more prevalent, and data grows yet larger, 

it will become more important for computing platforms to accommodate large-scale and unpredictable 
workloads. \NAME achieves this along three critical dimensions: security, scalability and cost-effectiveness.

\section{Availability}

\NAME is open source and is available at:
\begin{center}
{\tt https://github.com/yadudoc/cloud\_kotta}
\end{center}

\section*{Acknowledgments}
The authors thank Nandana Sengupta, Nathan Bartley, and Cha Chen for developing applications on \NAMENS.
This research was supported by grants from IBM for Computational Creativity, and a gift from Facebook.




\bibliographystyle{IEEEtran}
\bibliography{references}

\begin{thebibliography}{10}
\providecommand{\url}[1]{#1}
\csname url@samestyle\endcsname
\providecommand{\newblock}{\relax}
\providecommand{\bibinfo}[2]{#2}
\providecommand{\BIBentrySTDinterwordspacing}{\spaceskip=0pt\relax}
\providecommand{\BIBentryALTinterwordstretchfactor}{4}
\providecommand{\BIBentryALTinterwordspacing}{\spaceskip=\fontdimen2\font plus
\BIBentryALTinterwordstretchfactor\fontdimen3\font minus
  \fontdimen4\font\relax}
\providecommand{\BIBforeignlanguage}[2]{{%
\expandafter\ifx\csname l@#1\endcsname\relax
\typeout{** WARNING: IEEEtran.bst: No hyphenation pattern has been}%
\typeout{** loaded for the language `#1'. Using the pattern for}%
\typeout{** the default language instead.}%
\else
\language=\csname l@#1\endcsname
\fi
#2}}
\providecommand{\BIBdecl}{\relax}
\BIBdecl

\bibitem{Mikolov}
T.~Mikolov, I.~Sutskever, K.~Chen, G.~S. Corrado, and J.~Dean, ``Distributed
  representations of words and phrases and their compositionality,'' in
  \emph{Advances in neural information processing systems}, 2013, pp.
  3111--3119.

\bibitem{Rosen-Zvi}
M.~Rosen-Zvi, T.~Griffiths, M.~Steyvers, and P.~Smyth, ``The author-topic model
  for authors and documents,'' in \emph{Proceedings of the 20th conference on
  Uncertainty in artificial intelligence}.\hskip 1em plus 0.5em minus
  0.4em\relax AUAI Press, 2004, pp. 487--494.

\bibitem{Zhang}
J.~Zhang, A.~Gerow, J.~Altosaar, J.~Evans, and R.~J. So, ``Fast, flexible
  models for discovering topic correlation across weakly-related collections,''
  in \emph{Proceedings of Empirical Methods in Natural Language Processing},
  2015.

\bibitem{teplitskiy2015amplifying}
M.~Teplitskiy, G.~Lu, and E.~Duede, ``Amplifying the impact of open access:
  Wikipedia and the diffusion of science,'' \emph{arXiv preprint
  arXiv:1506.07608}, 2015.

\bibitem{shi2015weaving}
F.~Shi, J.~G. Foster, and J.~A. Evans, ``Weaving the fabric of science: Dynamic
  network models of science's unfolding structure,'' \emph{Social Networks},
  vol.~43, pp. 73--85, 2015.

\bibitem{gerow2015proposing}
A.~Gerow, B.~Lou, E.~Duede, and J.~Evans, ``Proposing ties in a dense
  hypergraph of academics,'' in \emph{International Conference on Social
  Informatics}.\hskip 1em plus 0.5em minus 0.4em\relax Springer, 2015, pp.
  209--226.

\bibitem{chard2015costworkloads}
R.~Chard, K.~Chard, K.~Bubendorfer, L.~Lacinski, R.~Madduri, and I.~Foster,
  ``Cost-aware elastic cloud provisioning for scientific workloads,'' in
  \emph{Proceedings of the 8th International Conference on Cloud Computing
  (CLOUD)}.\hskip 1em plus 0.5em minus 0.4em\relax IEEE, June 2015, pp.
  971--974.

\bibitem{chard2015cost}
------, ``Cost-aware cloud provisioning,'' in \emph{Proceedings of the 11th
  International Conference on e-Science}.\hskip 1em plus 0.5em minus
  0.4em\relax IEEE, August 2015, pp. 136--144.

\bibitem{chard2016profiling}
R.~Chard, K.~Chard, K.~Bubendorfer, A.~Rodriguez, R.~Madduri, and I.~Foster,
  ``An automated tool profiling service for the cloud,'' in \emph{Accepted to
  the 16th IEEE/ACM International Symposium on Cluster, Cloud and Grid
  Computing}, 2016.

\bibitem{s3_transfer_cost}
``{S3 Inter-region Transfer cost},'' \url{http://aws.amazon.com/s3/pricing/},
  web site. Accessed: May, 2016.

\bibitem{borner2011plug}
K.~B{\"o}rner, ``Plug-and-play macroscopes,'' \emph{Communications of the ACM},
  vol.~54, no.~3, pp. 60--69, 2011.

\bibitem{babuji2016secure}
Y.~N. Babuji, K.~Chard, A.~Gerow, and E.~Duede, ``A secure data enclave and
  analytics platform for social scientists,'' \emph{arXiv preprint
  arXiv:1610.03105}, 2016.

\bibitem{abramson14hybrid}
S.~Abramson, W.~Horka, and L.~Wisniewski, ``A hybrid cloud architecture for a
  social science research computing data center,'' in \emph{Proceedings of the
  34th International Conference on Distributed Computing Systems Workshops
  (ICDCSW)}, June 2014, pp. 45--50.

\bibitem{saleem14bigexcel}
M.~A. Saleem, B.~Varghese, and A.~Barker, ``Bigexcel: A web-based framework for
  exploring big data in social sciences,'' in \emph{Proceedings of the IEEE
  International Conference on Big Data (Big Data)}, Oct 2014, pp. 84--91.

\bibitem{zheng14capsules}
J.~Zeng, G.~Ruan, A.~Crowell, A.~Prakash, and B.~Plale, ``Cloud computing data
  capsules for non-consumptiveuse of texts,'' in \emph{Proceedings of the 5th
  ACM Workshop on Scientific Cloud Computing (ScienceCloud)}.\hskip 1em plus
  0.5em minus 0.4em\relax ACM, 2014, pp. 9--16.

\bibitem{dbgap}
M.~Mailman, M.~Feolo, Y.~Jin, M.~Kimura, K.~Tryka, R.~Bagoutdinov, L.~Hao,
  A.~Kiang, J.~Paschall, L.~Phan, N.~Popova, S.~Pretel, L.~Ziyabari, M.~Lee,
  Y.~Shao, Z.~Wang, K.~Sirotkin, M.~Ward, M.~Kholodov, K.~Zbicz, J.~Beck,
  M.~Kimelman, S.~Shevelev, D.~Preuss, E.~Yaschenko, A.~Graeff, J.~Ostell, and
  S.~Sherry, ``The {NCBI dbGaP} database of genotypes and phenotypes.''
  \emph{Nature Genetics}, vol.~39, no.~10, pp. 1181--1186, 2007.

\bibitem{ncdc}
``{National Climatic Data Center (NCDC)},'' \url{http://www.ncdc.noaa.gov/},
  web site. Accessed: May, 2016.

\bibitem{irods}
A.~Rajasekar, R.~Moore, C.-y. Hou, C.~A. Lee, R.~Marciano, A.~de~Torcy, M.~Wan,
  W.~Schroeder, S.-Y. Chen, L.~Gilbert, P.~Tooby, and B.~Zhu, \emph{iRODS
  Primer: Integrated Rule-Oriented Data System}.\hskip 1em plus 0.5em minus
  0.4em\relax Morgan and Claypool Publishers, 2010.

\bibitem{cloudman}
E.~Afgan, D.~Baker, N.~Coraor, H.~Goto, I.~M. Paul, K.~D. Makova,
  A.~Nekrutenko, and J.~Taylor, ``Harnessing cloud computing with galaxy
  cloud,'' \emph{Nature Biotechnology}, vol.~29, pp. 972--974, 2011.

\bibitem{starcluster}
``{StarCluster},'' \url{http://star.mit.edu/cluster/}, web site. Accessed: May,
  2016.

\bibitem{madduri2014globus}
R.~Madduri, K.~Chard, R.~Chard, L.~Lacinski, A.~Rodriguez, D.~Sulakhe,
  D.~Kelly, U.~Dave, and I.~Foster, ``The {Globus Galaxies} platform:
  delivering science gateways as a service,'' \emph{Concurrency and
  Computation: Practice and Experience}, vol.~27, no.~16, pp. 4344--4360, 2015.

\bibitem{albrecht12makeflow}
M.~Albrecht, P.~Donnelly, P.~Bui, and D.~Thain, ``Makeflow: A portable
  abstraction for data intensive computing on clusters, clouds, and grids,'' in
  \emph{Proceedings of the 1st ACM SIGMOD Workshop on Scalable Workflow
  Execution Engines and Technologies}.\hskip 1em plus 0.5em minus 0.4em\relax
  ACM, 2012, pp. 1:1--1:13.

\bibitem{wikinsdiehr07gateways}
N.~Wilkins-Diehr, ``Special issue: Science gateways—common community
  interfaces to grid resources,'' \emph{Concurrency and Computation: Practice
  and Experience}, vol.~19, no.~6, pp. 743--749, 2007.

\bibitem{liu13cybergis}
Y.~Liu, A.~Padmanabhan, and S.~Wang, ``Cybergis gateway for enabling data-rich
  geospatial research and education,'' in \emph{Proceedings of the IEEE
  International Conference on Cluster Computing (CLUSTER)}, Sept 2013.

\bibitem{stanzione11iplant}
D.~Stanzione, ``The iplant collaborative: Cyberinfrastructure to feed the
  world,'' \emph{Computer}, vol.~44, no.~11, pp. 44--52, Nov 2011.

\bibitem{wu11cloud}
W.~Wu, H.~Zhang, Z.~Li, and Y.~Mao, ``Creating a cloud-based life science
  gateway,'' in \emph{Proceedings of the 7th IEEE International Conference on
  e-Science}, Dec 2011, pp. 55--61.

\end{thebibliography}
%

\end{document}